# A Decision Support System for Public Research Organizations Participating in National Research Assessment Exercises[1]

Giovanni Abramo[*]

*National Research Council of Italy and Laboratory for Studies of Research and Technology Transfer at University of Rome "Tor Vergata" – Italy*

ADDRESS: Dipartimento di Ingegneria dell'Impresa, Università degli Studi di Roma "Tor Vergata", Via del Politecnico 1, 00133 Roma - ITALY, tel. and fax +39 06 72597362, abramo@disp.uniroma2.it

Ciriaco Andrea D'Angelo

*Laboratory for Studies of Research and Technology Transfer at University of Rome "Tor Vergata" – Italy*

ADDRESS: Dipartimento di Ingegneria dell'Impresa, Università degli Studi di Roma "Tor Vergata", Via del Politecnico 1, 00133 Roma - ITALY, tel. and fax +39 06 72597362, dangelo@disp.uniroma2.it

[1] Abramo, G., D'Angelo, C.A. (2009). A decision support system for public research organizations participating in national research assessment exercises. *Journal of the American Society for Information Science and Technology*, 60(10), 2095-2106. DOI: 10.1002/asi.21126

[*] **Corresponding author**

**Abstract**

We are witnessing a rapid trend towards the adoption of exercises for evaluation of national research systems, generally based on the peer review approach. They respond to two main needs: stimulating higher efficiency in research activities by public laboratories, and realizing better allocative efficiency in government funding of such institutions. However the peer review approach is typified by several limitations that raise doubts for the achievement of the ultimate objectives. In particular, subjectivity of judgment, which occurs during the step of selecting research outputs to be submitted for the evaluations, risks heavily distorting both the final ratings of the organizations evaluated and the ultimate funding they receive. These distortions become ever more relevant if the evaluation is limited to small samples of the scientific production of the research institutions. The objective of the current study is to propose a quantitative methodology based on bibliometric data that would provide a reliable support for the process of selecting the best products of a laboratory, and thus limit distortions. Benefits are twofold: single research institutions can maximize the probability of receiving a fair evaluation coherent with the real quality of their research. At the same time, broader adoptions of this approach could also provide strong advantages at the macroeconomic level, since it guarantees financial allocations based on the real value of the institutions under evaluation. In this study, the proposed methodology has been applied to the hard science sectors of the Italian university research system for the period 2004-2006.

# 1. Introduction

In recent years, governments of the more industrialized nations have favored a model of development based on the knowledge economy. As part of policies to sustain the competitiveness of industry, this choice includes and entails - above all - the reinforcement and the enhancement of public research infrastructure. However, policies targeted in this sense must take account of both increasing costs of research on the one hand and, at the same time, stringent budget limits. This situation has stimulated the adoption of resource allocation systems that have been initially based on criteria of quality, but are now moving to embed efficiency too. Towards such systems, many nations have begun to conduct national research assessments, intended to respond to three of their pressing needs: i) stimulation of better efficiency in research activities in public laboratories; ii) improvement in allocative efficiency in funding public research organizations by governments, and iii) reducing asymmetric information in the market for new knowledge, between supply (from universities, public research laboratories) and demand (by students, companies, etc.).

At a global level, there is a rapid diffusion of evaluation exercises for national research systems, typically based on "outcome control" logic and peer-review approach. The most significant experience is definitely that of the "Research Assessment Exercise" in Great Britain (RAE, 2008). The fifth edition of the assessment has just been concluded, aimed at assessing the quality profiles of all UK higher education institutions and using them for research funding decisions, with effect from 2009-10. This exercise is essentially based on evaluations by a panel of experts, who review a part of the scientific output of the research staff of each university. The RAE supports decisions for the allocation of not less than 25% of the total government funding destined for any university. Similar exercises can be found in other English-speaking nations, most prominent being Australia's Research Quality Framework (RQF, 2007) and New Zealand's Performance-Based Research Fund (PBRF, 2008). In the Netherlands, several evaluation exercises have been conducted, although the results are not used for allocative purposes. In 2006, Italy saw the conclusion of its first Triennial Research Evaluation (VTR), which was entrusted to a national Steering Committee for the Evaluation of Research (CIVR, 2006). The current government intends to allocate a growing portion of its university research funding (30% in 2011) on the basis of results from the national evaluation.

Scholars, policy makers, scientists and top managers of research institutions are increasingly involved in debates as to the strong and weak points of these exercises (Georghiou and Larédo, 2005). However there is ample consensus that the permanent adoption of these types of initiative is desirable, provided that their primary goal is to encourage and reward excellence of research in the tertiary education sector.

The literature on evaluation of research does not lack for methodological criticisms. In particular, the peer review approach appears afflicted by several limitations of subjectivity (Moxham and Anderson, 1992; Horrobin, 1990), which are seen in the selection of peer experts called to evaluate the single products; again in the work of the peers during evaluation of level of excellence; and also in the upstream process of choosing the research outputs to be submitted for evaluation, as selected by the single research institutions. The case of subjectivity during the step of selecting the products to be submitted can ultimately determine distorted ratings of excellence, with consequent inefficiency in allocation of resources among universities (where the evaluation is intended for these purposes). The distortion created by this limit becomes greater as the output of research subject to evaluation decreases. Yet the costs and times for national evaluation exercises would increase if, as a remedy, the number of products taken into consideration were increased. For this reason it seems of the highest priority to minimize the extent of the errors committed in the phase of selecting products to be evaluated, so that these result in being truly the best present in the scientific portfolios of the organizations being evaluated. However, a recent investigation (Abramo et al., 2008a) shows that the selection processes adopted by Italian universities for their participation in the VTR assessment are rather incoherent and dysfunctional, as analyzed by bibliometric criteria. Consequently, distortion in the overall result of the assessment exercise may be partially due to the ineffective initial selections made by a number of universities, which evidently do not possess sufficient instruments for the selection itself or are particularly subject to rank and seniority pressures.

The objective of this work is to propose a quantitative methodology based on bibliometric data that contributes to solving problems of subjectivity, by providing a reliable support for the process of selecting the best products of an organization and thus limiting the distortions noted above. In the "hard sciences" and in some sectors of the economic sciences, publications in international scientific journals constitute the primary codified form of research results, most often (if not exclusively) adopted by research scientists. In these scientific sectors the quality of a research output is indirectly guaranteed by the peer review process it must pass in order to be published, and further judged by the citations that the publication then receives. In these cases, bibliometric impact indicators can represent a proxy value for measurement of the quality of a work (Oppenheim, 1997; Oppenheim and Norris, 2003; Rinia et al., 1998) and thus support the process of identifying the best products present in the scientific portfolio of the organization[2].

The methodology proposed by the authors was applied, in this study, to the Italian national university system, for the 2004-2006 triennium. It is based on the scientific products listed in the Thomson Reuters' "Web of Science" (WoS) for each of the 82 universities that compose the Italian national system, and on the evaluation of the bibliometric positioning of each publication (on the basis of its citations received and the impact factor of the publishing journal), with respect to the totality of products listed in the WoS reference scientific sector.

The proposed approach can provide a useful support for the participation of single research institutions in the national evaluation exercises, in maximizing the probability that the institution will receive an evaluation that is not below the real quality of its research. Broader adoption of such approaches can also determine significant advantages at the macroeconomic level. By reducing the distortions noted in the preceding paragraphs, it guarantees a more effective government allocation of public funds, in which funding is coherent with the true value of the organizations evaluated. On the one hand, a single public research institution would wish to make an efficacious choice of its research output, while at the same time it can be seen that government would also desire the same situation. The advantages of bibliometric measurements in supporting peer-review are increasingly acknowledged at government level, as shown by the changes that are occurring in a number of national research assessment systems.

---

[2] Literature on this issue stresses the intrinsic limits of the use of impact indicators to represent the quality of a given publication. The journal impact factor in particular, representing the mean value of citations received for articles published in a given journal, offers a poor predictor of the quality of single articles. The actual counting of citations is also a proxy that is not without imperfection: a given article could be cited not because it is seen as particularly useful in a given argument, but rather to be criticized or even refuted. In spite of this situation, quantitative studies by bibliometrists agree in acknowledging a strong correlation between the absolute quality of a work (measured according to judgments from peers) and the relative impact indicators, and this justifies their continued use.

In future, the UK is moving to a Research Excellence Framework (REF) that will be primarily, if not exclusively, metrics-based for the hard sciences[3]. In Australia, the RQF has been abandoned for the new Excellence in Research for Australia (ERA). As with the REF, this will be predominantly a metrics-based exercise[4]. Bibliometric approaches are in fact economical, non invasive, and simple to implement; they permit updates and rapid inter-temporal comparisons; are based on more objective[5] quantitative data; are capable of examining a higher representation of the universe under investigation and adapt well to international comparisons. Above all the bibliometric approach may allow productivity measures[6] and thus the integration of quality with efficiency criteria in research assessment (Abramo et al., 2008b). We underline that bibliometrics should be considered as a complement rather than a substitute for peer-review.

The proposed decision support system can be applied in any context, where a selection of products is required to participate in research assessment exercises. The Italian VTR has been derived from the British RAE, which is similar to the Australian RQF. They are all based on peer-review of a sample of research products, thus implying selection by research institutions under evaluation.

The current study is organized as follows: the next section will highlight the importance of the theme taken on, through a critical analysis of the methodologies of various national evaluation exercises and an examination of the Italian case in particular. The third section presents the support system proposed by authors, in terms its methodology and the dataset to be used. The fourth section presents the results of applying the proposed methodology to the Italian university system for the period 2004-2006. The concluding section gives the conclusions and recommendations of the authors.

---

[3] Details can be found at: http://www.hefce.ac.uk/research/ref/
[4] Details can be found at: http://www.arc.gov.au/era/default.htm
[5] Generally, there is a pattern of good correspondence between bibliometric indicators and peer judgements, which has sometimes led to pointing to bibliometric indicators as objective measures that contrast with the subjective character of peer review. However Weingart (2003) reminded that bibliometric indicators themselves are based in part on peer decisions, since journal articles embody the peer evaluations that led to their acceptance for publication.
[6] Once the most obstinate obstacle confronting the bibliometrician is overcome. This is the accurate attribution of articles to their true authors and institutions, due to homonyms in author names and to the failure of the source databases to indicate the individual connections between the authors listed and the accompanying lists of research institutions. This obstacle has, until now, severely limited the full use of bibliometric evaluations at a national scale.

## 2. The process of selecting outputs by public research organizations: a retrospective analysis of the Italian 2001-2003 assessment exercise

National exercises in evaluation of research systems are generally concerned with quality and impact of scientific production, but also dimensions such as level of internationalization of research activity, results in technological transfer, capacity to sustain competitiveness in the territories, capacity to attract external funding, etc. However, the excellence of scientific production is invariably the component given the most weight. In the Britain's RAE, the weight of this dimension varies from discipline to discipline, between a minimum of 50% and a maximum of 70%. In the Italian assessment, for the final evaluation profile of the institutions over all disciplines, the excellence of scientific production invariably assumed a weight of 60%.

It is therefore clear that the principal focus of national exercises concerns evaluations of the products of research, for which each national agency charged with conducting an assessment first sees to the establishment of panels of experts, for every discipline in which it intends to conduct an evaluation. Universities and public research laboratories (termed together as "public research organizations"; abbreviated "PROs") are requested to select a certain number of products to be submitted to the panels, in numbers generally proportional to the number of research personnel employed in the various discipline of activity for the PRO. The products selected are thus taken in by the panels and then redistributed to the experts pertaining to the scientific sector for each product. Typically, each research output is then evaluated by at least two experts, originating from national or international spheres. The experts are requested to give qualitative judgments on a scale of possible ratings, typically graduated in 4 levels. The judgments by these experts are then re-examined by overall panels and aggregated by discipline for all PROs. In this manner the exercise arrives at a rating expressing the research profile of each organization under evaluation, for each discipline.

The recently conducted Italian evaluation exercise the analysis concerned 14 disciplines, conforming to the national classifications of university research personnel, plus 6 cross-sector disciplines, with each discipline entrusted to its own expert panel. The assessment system was designed to evaluate research and development carried out by all PROs, through obligatory participation of universities and optional participation from public research laboratories. A total of 102 PROs were evaluated. The overall total of products evaluated for 2001-2003 was 17,329, of which 13,374 were produced by universities. PROs were asked to autonomously select and submit research outputs[7] to the panels. The number of research products submitted was to equal to 25% of the number of full-time-equivalent researchers working in the university (50% for public research laboratories) in the period under observation. Within this overall restriction on selections, it was not compulsory that the ratio be respected for each discipline, leaving the PROs the possibility to privilege some sectors (representing them with a greater number of products) with respect to others (representing them with a lower number of selections).

---

[7] "Research output" was understood as a result of research activity codified in various potential forms: articles in journals, monographs, contributions to monographs, patents, industrial prototypes, artworks, etc. In reality, for most disciplines, universities made scientific publications their almost exclusive selection.

Each organization was free to adopt its own strategies and methods for making selections but, in general, a tendency emerged of drawing on small scale peer review methods for comparative evaluation and selection of "best" products, which were then submitted to the national panels. These procedures resulted as costly, inefficient and ultimately lacking efficacy, as demonstrated in the study by Abramo et al. (2008a). Through a bibliometric analysis of the publications selected in the hard sciences (Mathematics and computer science, Physics, Chemistry, Earth science, Biology, Medicine, Agricultural and veterinary science, Industrial and information engineering), the study showed that universities, in most cases, did not identify and/or present their supposedly best publications. For three of these disciplines (Agricultural and veterinary science, Industrial and information engineering and Mathematics and computer science) the universities selected a sample of articles that, in 30%, 27% and 25% of the respective sectors, presented a normalized impact factor (IF)[8] that was below the median IF of the total Italian university publication portfolio for the discipline. In these disciplines, and in Earth science, Physics and Chemistry, there were actually universities that presented only publications with an IF lower than the median of the IF distribution of their complete publications in each discipline. As an example, in Industrial and information engineering, the University of Rome "Tor Vergata" presented 14 international journal articles. Of these, only 3 appeared among the 14 top IF-ranked articles that had been produced by the pertinent departments. In all probability, both social ("parochialism") and technical factors (the real difficulty of comparing articles from diverse sectors), may have compromised the effectiveness of the selection process[9].

---

[8] The impact factor is an index of the impact of a journal, traceable to the citations received by the articles that it publishes. This index can be normalized with respect to the mean of all the journals dealing with the same WoS discipline, thus enabling comparison between journals dealing with different disciplines.

The distortions in input to the Italian assessment exercise are clearly a situation stemming from the relative immaturity of evaluation in the nation: for many research scientists, department directors and deans, the recent VTR was the first experience with an evaluation. Furthermore, the low quantity of product to be submitted rendered the selection process more complicated. We could imagine a representative case in which a university is choosing only one product from four publications in medical science: choosing from among one publication in the cardiology sector, one in dermatology, one in orthopedic medicine and one in neurosciences. An efficacious selection from such diverse material could be greatly assisted by bibliometrics. But many universities do not have access to bibliometric data bases such as WoS by Thomson Reuters, Scopus by Elsevier or other similar indexes, and so for the VTR they had to resort to internal peer review: it is difficult to imagine that anyone could possess the competencies, for all four disciplines of our example, to be able to choose the best product, with reasoned logic. This type of problem does not exist downstream, once the selected products arrive at the national panels for evaluation. At this level, in fact, experts evaluate only products from their field of specialization: cardiologists are asked to assess only the quality of products in cardiology; dermatologists in dermatology, and so on.

For the universities that do draw on the Web of Science or analogous bibliometric databases for assistance, the selection of products in the hard sciences could result as more manageable and efficacious, provided it is carried out with the necessary attention to detail. A simple comparison of impact indicators (citations per article, or IF of the relative journal) to identify the best publications could lead to erroneous choices, due to at least two causes of distortion: i) the different numbers of journals listed in the bibliometric data bases, per scientific sector; ii) the differences in production intensity, or "fertility" of production, inherent to the different scientific sectors under evaluation. These causes would lead, with publications of equal quality, to differences in probability for their citation depending on the scientific sectors in which they fall. Foresight would be necessary, before conducting a comparison, so that the selection process would provide for normalization of the impact indicators relative to the mean value of each sector, a datum which is generally available from the more sophisticated data bases[10]. In this manner it is possible to avoid the distortions in comparing among publications originating in different scientific sectors.

---

[9] Its bottom-rank in the VTR, among universities of similar size, contrasts with the top 30 percentile in metrics-based excellence ranking.

A further item of information that would surely be useful to decision-makers within a PRO would be the relative positioning, for impact, of each publication in its scientific sector with respect to all the competition (meaning articles authored by other PROs). We can expect that the peers on national panels, selected for their competencies in specific fields, will use comparative logic, taking into consideration their knowledge of other products, assigned to them for evaluation, as they are forming their judgments. For this reason, decision-makers within an individual PRO could be interested in knowing, relative to a single product, not only its quality with respect to others produced by the same PRO, but also relative to competitive products in the same scientific sector from other organizations. Continuing our example, the article on cardiology could have an impact index slightly less than that of the dermatology article, but result first among the competitor publications from other PROs, while that in dermatology could place last in its analogous classification.

---

[10] The desired situation would be that the impact indicator distributions in each scientific discipline are not highly skewed, because other statistics, such as the median, are not provided.

The method proposed in the present work offers decision-makers within any PRO a support system for decisions that responds to all of the above needs and considerations.

## 3. The proposed methodology

The support system, developed here for the use of PROs participating in any national research assessment, will be illustrated with reference to the Italian case, in particular to the case of Italian universities. The intention is that its application can be easily adapted to other public research laboratories and national systems.

The data set of scientific products for the study was derived directly from the Observatory of Public Research (ORP, 2008), created and developed by the authors as part of their work with the Laboratory for Studies of Research and Technology Transfer at the University of Rome "Tor Vergata". The ORP is a data base derived under license from the Thomson Reuters Web of Science, which indexes the scientific production, beginning in 2001, for all Italian PROs (82 universities, 76 public research laboratories, 200 research hospitals). The field of observation used to illustrate the methodology covers the 2004-2006 triennium and is limited to the hard sciences, particularly to 8 of the total 14 universities disciplines (UD) that compose the Italian university system. These 8 UD are Mathematics and computer science, Physics, Chemistry, Earth science, Biology, Medicine, Agricultural and veterinary science, Industrial and information engineering. In the Italian case, these sectors include 60% of total national university research personnel. In addition, for these UD, 95% of the total research outputs submitted by universities for the Italian VTR were also research articles indexed in the ORP, with a minimum of 89% in Industrial and information engineering and a maximum of 97% in Medicine (Table 1).

[Table 1]

For the purposes of the proposed study, the census included only publications classified in the indexes as "article" or "review". In the period of observation, there were 122,494 of these publications, of which 95% were articles. For the Web of Science, each of these is classified as to its scientific sector. The authors grouped the WoS hard science sectors into the 8 UD under examination (see Annex). The quality of each publication was measured in two dimensions: the first referring to the journal and the second referring directly to the article. For this latter dimension the measure was conducted using two methods. Thus the three indicators are defined as follows:

i. *Journal Impact Ranking* (JIR). Impact Factor ranking of the journal, measured on a 0 – 100 percentile scale according to the Impact Factor distribution of the journals falling in the same WoS sector. A value of 90 indicates that 90% of the journals falling in the same sector have lower impact factor than the one at stake.

ii. *Article Impact Ranking* (AIR). Citation ranking of an article, measured on a 0 – 100 scale according to the citation distribution of the articles of the same year falling in the same WoS sector. A value of 90 indicates that 90% of the articles of the same year falling in the same sector have a lower number of citations than the article under consideration.

iii. *Article Impact Index* (AII). Number of citations of an article divided by the average number of citations of all articles of the same year, falling in the same WoS sector. A value of 1.40 indicates that the article was cited 40% more often than the average.

Providing more than one quality indicator is deemed necessary because they may be more or less meaningful for different scientific disciplines and ages of publication. Several quality indicators were used for their potential to provide more meaningful results with respect to the diverse products of different scientific sectors and different "ages" of publications.

The operating procedure was articulated in 4 steps:

i. Census of all scientific production of the PROs for the triennium under observation. This step required that the authors identify and reconcile the different naming[11] with which the authors indicated their organizational affiliation in the publications. In order to establish consistency, given the triennium in consideration, this task required over 20,000 reconciliation rules, since each of the hundreds of Italian research organizations is cited in a wide variety of forms by the authors of the various publications[12].

---

[11] A specific test conducted by the authors revealed that the system of providing author addresses on the Web of Science leads to numerous unacceptable errors. For example, the WoS indicates the acronym "IRCCS" as representing a specific Italian research organization. In reality, it stands for a very generic title that the Italian Ministry of Health gives to over 40 different public and private hospitals which carry out research.

[12] For example, the address of the University of Rome "Tor Vergata" is given in 262 distinct forms among the publications included in the data set.

ii. Measurement of the quality of each publication, on the basis of the three bibliometric indicators noted above and comparing to other publications from Italy falling within the same WoS sector.

iii. Assigning each publication to the UD to which it pertains. WoS assigns each publication to one scientific sector. With the help of discipline specialists we have created a correspondence between WoS sectors and each UD (see Annex).

iv. For each UD, identification of the best publications of the PRO under evaluation: for each of the three indicators, a ranking of the publications of the PROs is developed and the best publications are selected in a recursive manner, until the intersection of the three selected sets contains a number of publications at least equal to the number of products that would be required for presentation to a hypothetical national evaluation exercise.

## 4. Results

To illustrate how the proposed support system works we will refer to the case of a specific university: the University of Milan, which is the fourth-ranking university in Italy in terms of number of research staff, after the University or Rome "La Sapienza", University of Bologna and University of Naples "Federico II". Our intention is to demonstrate that the proposed methodology will be sufficiently robust to permit its manageable and effective application to other types of PROs, irrespective of type or size. For a given PRO, the complexity of selecting products depends on the size of its total portfolio of research outputs and the variety of scientific sectors represented. The proposed methodology will thus be of significant assistance to medium and large-sized organizations, active in a number of sectors.

Supposing that, in a hypothetical national evaluation exercise covering a given period, each PRO under consideration is requested to present one product for every 4 scientists on staff in each discipline in which the organization is active. (This is the selection actually requested by the recent Italian evaluation of university research.) Table 2 (columns 1 to 4) presents data concerning the situation that would confront the University of Milan. Given its average of 1,670 researchers on staff over the 2004-2006 triennium, in the 8 disciplines under consideration[13], the university must select 418 products. The choices to be made represent slightly more than 5% of the entire scientific production of the university in the UD under observation, with a low representation of 1% from Industrial and information engineering[14] and a maximum of 14.5% from Agricultural and veterinary science. This first observation of the results is particularly important because:

- the sample of products submitted for evaluation is truly limited: under the rules of the preceding Italian assessment, one product of every 20 produced at the university is selected;
- the representativeness of the sample varies from discipline to discipline, because each is characterized by a different level of scientific productivity.

---

[13] In the Italian university system, each scientist is classified as pertaining to a scientific sector, which in turn pertains to a particular UD.
[14] This discipline represents an anomaly for the University of Milan: although it has only 20 research personnel, we recorded 585 publications. These numerous publications very likely originate in large part from researchers on staff in other disciplines. The apparent "imbalance" can occur because the classification of products per WoS discipline is based on the product itself, and not on the disciplinary affiliation of the authors, given according to the Italian national university classification.

From among the publications listed for each UD, following the procedure described in Section 3 of this study, the three sets of best publications are then identified based on their national positioning for each of the three indicators of quality. Table 2, column 6 presents the case for the University of Milan, giving the number of best publications per UD, net of the double counting in the three sets (the same publication can appear in more than one set). The entire selection process identifies 2,020 publications, representing 25.2% of the university's total production for these UD, with a minimum of 16.2% from Industrial and information engineering and a maximum of 42.1% from Agricultural and veterinary science. The last column of Table 2 shows the ratio of number of publications furnished to the decision-makers of each UD relative to the number of publications to be submitted to eventual evaluation.

[Table 2]

The university body elected to make the final selection of products for submission to the national panels thus receives a list of publications indicating the criteria of selection (JIR, AIR or AII) for each publication, and their relative ranking[15]. As a result of the structure of the procedure, this list contains the intersection set (column 6 of Table 3) consisting of those publications that simultaneously appear in all three sets, as extracted for each indicator. The responsible body at the university can concentrate their final selection on this intersection set of publications, or exercise their discretion to opt for others. This flexibility is useful because the bibliometric indicators may be more or less meaningful for different scientific sectors and "ages" of publication. For example in

---

[15] The panels obviously also receive also the essential bibliometric information on the publications (title, authors, journal and other publication details).

Biology, the citations received per article could certainly be seen as a more trustworthy proxy of the real impact of the article on the international scientific community, as compared to the impact factor of the publishing journal. However, this may not always hold true: for example, in the case of Mathematics, the life cycle of citations of an article is particularly long and data monitored at only 2 to 3 years from the date of publication may not be a reliable indicator of the real value of the work. An additional observation is that, in all the UD, the distribution of citations for recent works often reflects chance rather than the real quality of a publication[16]. For these various reasons, the evaluation of the publications of a particular discipline could draw on various indicators, giving each a suitable weighting in keeping with the characteristics of the specific discipline. Table 3 (columns 3 to 5) presents the number of publications selected for the University of Milan on the basis of each indicator[17].

[Table 3]

The university body responsible for the final selection of products to submit to the national panels of peers can decide whether to emphasize (or even use exclusively) one of the three ranking criteria in preference to the others, for example preferring to consider impact factor over citations. Or, with respect to citations, it can decide to give greater weight to AIR over AII, but still select publications with a particularly high level of AII (which expresses the ratio of citations received relative to the sector mean), independently of the value of AIR. Or, as initially suggested above, it can simply select publications that appertain to the intersection of the three top-ranked sets for each indicator (Table 3, column 6).

---

[16] A work published in January 2006 could have more citations that one published in December of the same year, but not necessarily because it is qualitatively better.
[17] In some cases the number of publications for the three indicators is not identical because of multiple counts from publications of equal rank at the minimum entry threshold.

The proposed methodology could provide a further support to the university body responsible for the final selection of products, as seen from Table 4. This gives a type of summary view indicating, for each discipline, the average values of the three indicators both for the total 2,020 publications and for 425 publications belonging to the intersection of the 3 sets. While not presuming to represent the difference in excellence between the disciplines under consideration, this data could suggest favoring one discipline over another in the final selection of products, provided that the university's evaluation body has a certain margin of flexibility. As a case in point, in the last evaluation exercise for Italy, the CIVR did not indicate that the ratio of product selection per scientist had to be observed for each discipline within every university, but rather only at the aggregated level of the entire university. In the current example, the University of Milan could have decided to present a greater number of products from discipline 4 (Earth sciences) and 6 (Medicine) and a lesser number from the other disciplines, with the aim of maximizing its overall final rating.

[Table 4]

5. **Distribution of best products**

One of the requisites of a support system for the process of selecting the best scientific products of a university is the necessity of robust results, including impartiality with respect to the age of the products and their sectorial classification. The proposed system, owing to its method, guarantees that the publications identified are not concentrated in certain years of production or in certain scientific sectors of the relevant discipline.

We verify this characteristic with the example of the University of Milan, first considering the age variable. As illustrated by the data in Table 5, there is no significant polarization: the number of publications identified for each year is consistently around one third of the total, owing to the selection process being based on bibliometric indicators that are normalized with respect to the time variable. The variations, if they occur, are truly due to variations in scientific productivity registered in a given discipline, year for year. This robustness is particularly important if one considers that, in peer review selection processes, it is sometimes erroneously argued that recent works should be favored because, at the moment of evaluation, they are said to be closer to the knowledge frontier.

[Table 5]

The same robustness can be observed in the sectorial classification of products identified through the proposed procedure. Table 6 shows the details for 361 publications selected for the Biology discipline at the University of Milan: these are distributed over 20 WoS sectors. The sector with the most representation is Biochemistry and molecular biology, with 92 publications identified, representing 25.5% of the total of 'best publications". Meanwhile, this same sector achieves 24.2% of the overall scientific production in the Biology discipline. In general, as in this particular example, there is a strong correlation (0.97) between the third and fifth columns of Table 6, indicating that the sets of publications identified with the proposed procedure is strongly representative of the sectorial specializations present at the University.

[Table 6]

In this regard it can also be useful to compare the relative results of two universities. Table 7 presents the detail for the distribution of the best publications selected in the Biology discipline for the University of Milan and University of Rome "Tor Vergata". First, as for the University of Milan (Table 6), the validity of the publications as representative of sectorial specializations was also confirmed for the University of Rome "Tor Vergata" (correlation index 0.93). Table 7 then shows that "Biochemistry & molecular biology" is the WoS sector that is best represented for both universities. The second most frequent sector differs between the University of Milan (Cell biology) and "Tor Vergata" (Biology). Further, at the University of Milan, the products selected for "Plant sciences" and "Marine & freshwater biology" represent 10% of the total compared to 0% at Tor Vergata. The differences between the results for these two universities show that the proposed methodology guarantees the representation of the true specificity of each research organization, in terms of the specialization by sector in every discipline.

[Table 7]

## 6. Conclusions

This work presents a bibliometric support system for the selection of publications, to assist universities and other public research laboratories in the selection of research outputs that they submit for participation in national evaluation exercises.

Such exercises are increasingly adopted in various nations and they typically draw on the peer review approach. Being based on judgments by individuals, this approach presents evident risks. This work has concentrated on responding to risks arising from subjectivity in judgments by those responsible for each organization's internal choices of products to be submitted for evaluation. A retrospective analysis conducted by Abramo et al. (2008a), on the results of the first national evaluation of the Italian public research system, showed that the degree of error in this process is particularly high. Such error is caused by both social (parochialism) and technical factors (limited use of bibliometric selection methods). The support system for product selection developed in this study permits minimizing errors determined by technical limits, meaning the limits that organizations encounter in attempting to compare the real quality of research products among different sectors of the same discipline, and then choose those with greater chance of receiving positive judgments. Further, the presentation of quantitative and objective data of good quality could assist in reducing the margin for maneuvers by those decision-makers that are inclined to parochialism. Errors in product selection penalize not only the organizations that commit them, but also risk compromising the overall aims for which the evaluation exercise was actually conceived: to stimulate greater production efficiency, support the efficient allocation of financial resources and reduce asymmetric information between supply of and demand for new knowledge. The evaluation exercise may provide a measure, seemingly an accurate one, but if this measure concerns a body of research which does not exactly correspond to that which should be measured, then the policy interventions that follow will certainly not produce the desired effects. It could be that a significant portion of research funding is allocated on the basis of distorted ratings of excellence caused by errors in the input selection of best products, committed by the organizations being evaluated. The expected increase in efficiency will not take place. In addition, with small portions of research output being submitted to evaluation then the problem of error becomes greater: in the case of the first Italian national assessment, the University of Milan was permitted to present a product set that represented only 5% of its total scientific production, including a low of 3.4% for research in medicine.

The support system for the process of selecting product, as presented in this work, attempts to reduce such distortions through the use of quantitative, objective bibliometric data. The method is based on the census of the entire national scientific production in the hard sciences, and thus on comparison of every single publication with all the others of the same Web of Science sector. It furnishes each research organization with a subset of publications from which they can make the final selection for submission to the national evaluation panels. As many specialists in evaluation of research agree, bibliometrics should be combined with peer evaluation, as it can often provide a valid support for improving and optimizing assessments: the proposed system does not substitute the work of decision-makers within the PRO. The support system facilitates their task by winnowing, from all the products present in the portfolio of a PRO, those that have already established a significant impact on the scientific community of reference, demonstrated by bibliometric data. Although the system can be applied to all scientific sectors it results as most efficacious for the disciplines where scientific publications are the primary codified form of new knowledge, which is especially the case for research in the hard sciences and some sectors of economics. The method is less valid for arts and humanities.

The simulations conducted demonstrate that the system is robust. It identifies a set of publications that shows no polarization with respect to their "age" and which well represents the sectorial specializations of universities in each discipline.

Since the system is based on the query of pre-existing bibliometric data bases, it is not invasive: it does not require contact with individual authors to gather products for submission to potential selection. This aspect avoids opportunity costs for research personnel, both tangible and psychological, including natural resistance to carrying out an activity other than research. Such opportunity costs are not at all negligible in national exercises for evaluation of public research. The method also offers rapid implementation and reliability, guaranteed by the use of quantitative, standardized, homogenous and objective data.

Public research organizations will not find that direct recourse to the internet search version of bibliometric data bases, such as the Thomson Reuters Web of Science, can alone constitute a reliable solution. This is both because the WoS internet system for indicating addresses leads to a very high error rate in identifying the home institutions of authors, and because only the total census of national scientific production permits the correct evaluation of the relative quality of a given article within its pertinent WoS sector.

The authors hope for a rapid and generalized adoption of this approach, especially in nations where the evaluation of research is taking its first timid steps, such as in Italy. In fact the proposed methodology is not country-specific. It can be easily adapted to other national contexts. However, the authors do already recognize the need for a further consideration. Public research organizations could actually select their true best output, perhaps with the assistance of the proposed approach, and then receive evaluation and funding in keeping with the quality of their work. However this would still not achieve the desired macroeconomic effect if internal redistribution of government resources within each PRO does not follow a consistent logic: the desired effects of national evaluation systems for research can result only if a "funds for quality" rule is considered together with other strategic issues, at all levels of decision-making. Again, bibliometrics could provide a useful support to public research organizations by assisting in-house panels of peer reviewers to identify best scientists or research groups, and then allow them to distribute resources, within the discipline considered strategic, also on the basis of merit. The authors have worked for some time on developing a useful decision support system for this situation and foresee the early presentation of a further proposal, once again efficacious and robust.

# ANNEX – List of WoS scientific sectors and corresponding university disciplines (UD)

| WoS sectors | UD[18] | WoS sectors | UD |
|---|---|---|---|
| Acoustics | 2 | Ecology | 5 |
| Agricultural economics & policy | 7 | Electrochemistry | 3 |
| Agricultural engineering | 7 | Emergency medicine | 6 |
| Agriculture. dairy & animal science | 7 | Endocrinology & metabolism | 6 |
| Agriculture. multidisciplinary | 7 | Energy & fuels | 8 |
| Agronomy | 7 | Engineering. aerospace | 8 |
| Allergy | 6 | Engineering. biomedical | 8 |
| Anatomy & morphology | 5 | Engineering. chemical | 8 |
| Andrology | 6 | Engineering. electrical & electronic | 8 |
| Anesthesiology | 6 | Engineering. environmental | 8 |
| Anthropology | 5 | Engineering. industrial | 8 |
| Astronomy & astrophysics | 2 | Engineering. manufacturing | 8 |
| Automation & control systems | 8 | Engineering. marine | 8 |
| Behavioral sciences | 6 | Engineering. mechanical | 8 |
| Biochemical research methods | 5 | Engineering. multidisciplinary | 8 |
| Biochemistry & molecular biology | 5 | Engineering. ocean | 4; 8 |
| Biodiversity conservation | 5 | Engineering. petroleum | 8 |
| Biology | 5 | Entomology | 7 |
| Biophysics | 5 | Environmental sciences | 4; 8 |
| Biotechnology & applied microbiology | 5 | Ergonomics | 8 |
| Cardiac & cardiovascular systems | 6 | Evolutionary biology | 5 |
| Cell biology | 5 | Fisheries | 7 |
| Chemistry. analytical | 3 | Food science & technology | 7 |
| Chemistry. applied | 3 | Forestry | 7 |
| Chemistry. inorganic & nuclear | 3 | Gastroenterology & hepatology | 6 |
| Chemistry. medicinal | 3 | Genetics & heredity | 6 |
| Chemistry. multidisciplinary | 3 | Geochemistry & geophysics | 4 |
| Chemistry. organic | 3 | Geography. physical | 4 |
| Chemistry. physical | 3 | Geology | 4 |
| Clinical neurology | 6 | Geosciences. multidisciplinary | 4 |
| Computer science. artificial intelligence | 8 | Geriatrics & gerontology | 6 |
| Computer science. cybernetics | 8 | Gerontology | 6 |
| Computer science. hardware & architecture | 8 | Health care sciences & services | 6 |
| Computer science. information systems | 1; 8 | Health policy & services | 6 |
| Computer science. interdisciplinary applications | 8 | Hematology | 6 |
| Computer science. software engineering | 8 | Horticulture | 7 |
| Computer science. theory & methods | 1; 8 | Imaging science & photographic technology | 8 |
| Critical care medicine | 6 | Immunology | 6 |
| Crystallography | 4 | Infectious diseases | 6 |
| Dentistry. oral surgery & medicine | 6 | Information science & library science | 8 |
| Dermatology | 6 | Instruments & instrumentation | 8 |
| Developmental biology | 5 | Integrative & complementary medicine | 6 |

---

[18] This table indicates the UD as number codes: 1, Mathematics and computer science; 2, Physics; 3, Chemistry; 4, Earth science; 5, Biology; 6, Medicine; 7, Agricultural and veterinary science; 8, Industrial and information engineering.

| WoS sectors | UD | WoS sectors | UD |
|---|---|---|---|
| Limnology | 4 | Paleontology | 4 |
| Marine & freshwater biology | 5 | Parasitology | 6 |
| Materials science, biomaterials | 8 | Pathology | 6 |
| Materials science, ceramics | 8 | Pediatrics | 6 |
| Materials science, characterization & testing | 8 | Peripheral vascular disease | 6 |
| Materials science, coatings & films | 8 | Pharmacology & pharmacy | 6 |
| Materials science, composites | 8 | Physics, applied | 2 |
| Materials science, multidisciplinary | 8 | Physics, atomic, molecular & chemical | 2; 3 |
| Materials science, paper & wood | 8 | Physics, condensed matter | 2 |
| Materials science, textiles | 8 | Physics, fluids & plasmas | 2 |
| Mathematical & computational biology | 5 | Physics, mathematical | 1; 2 |
| Mathematics | 1 | Physics, multidisciplinary | 2 |
| Mathematics, applied | 1 | Physics, nuclear | 2 |
| Mathematics, interdisciplinary applications | 1 | Physics, particles & fields | 2 |
| Mechanics | 2 | Physiology | 5 |
| Medical ethics | 6 | Plant sciences | 5 |
| Medical informatics | 8 | Polymer science | 3 |
| Medical laboratory technology | 6 | Psychiatry | 6 |
| Medicine, general & internal | 6 | Psychology | 6 |
| Medicine, legal | 6 | Public, environmental & occupational health | 6 |
| Medicine, research & experimental | 6 | Radiology, nuclear medicine & medical imaging | 6 |
| Metallurgy & metallurgical engineering | 8 | Rehabilitation | 6 |
| Meteorology & atmospheric sciences | 4 | Remote sensing | 8 |
| Microbiology | 5 | Reproductive biology | 5 |
| Microscopy | 8 | Respiratory system | 6 |
| Mineralogy | 4 | Rheumatology | 6 |
| Mining & mineral processing | 8 | Robotics | 8 |
| Mycology | 7 | Soil science | 7 |
| Nanoscience & nanotechnology | 2; 3; 5; 8 | Spectroscopy | 2 |
| Neuroimaging | 6 | Sport sciences | 6 |
| Neurosciences | 6 | Statistics & probability | 1 |
| Nuclear science & technology | 8 | Substance abuse | 6 |
| Nursing | 6 | Surgery | 6 |
| Nutrition & dietetics | 6 | Telecommunications | 8 |
| Obstetrics & gynecology | 6 | Thermodynamics | 2 |
| Oceanography | 4 | Toxicology | 6 |
| Oncology | 6 | Transplantation | 6 |
| Operations research & management science | 1 | Tropical medicine | 6 |
| Ophthalmology | 6 | Urology & nephrology | 6 |
| Optics | 2 | Veterinary sciences | 7 |
| Ornithology | 5 | Virology | 6 |
| Orthopedics | 6 | Water resources | 4 |
| Otorhinolaryngology | 6 | Zoology | 5 |